\documentclass[twocolumn,superscriptaddress,showpacs,showkeys,titlepage,floatfix,balancelastpage,a4paper,10pt]{revtex4-1}
\usepackage{amsfonts,amssymb,amsmath,latexsym}
\usepackage{graphicx}
\usepackage{xcolor}
\usepackage{setspace}
\usepackage{setspace}
\usepackage{graphicx}
\usepackage{multirow}
\usepackage{mathtools}
\usepackage{tabularx}
 \RequirePackage{verbatim}
\newcommand{\bu}{\mathbf{u}}
\newcommand{\bv}{\mathbf{v}}
\newcommand{\bV}{\mathbf{V}}
\newcommand{\bW}{\mathbf{W}}
\newcommand{\bR}{\mathbf{R}}

\newcommand{\lcpq}{Laboratoire de Chimie et Physique Quantiques, IRSAMC, CNRS, Universit\'e de Toulouse, UPS, France}
\newcommand{\unibo}{Universit\`a di Bologna, Bologna, Italy}
\newcommand{\etsf}{European Theoretical Spectroscopy Facility (ETSF)}

\begin{document}
\title{Clifford boundary conditions: a simple direct-sum evaluation of Madelung constants}

\author{Nicolas Tavernier}
\affiliation{\lcpq}
\author{Gian Luigi Bendazzoli}
\affiliation{\unibo}
\author{V\'eronique Brumas}
\affiliation{\lcpq}
\author{Stefano Evangelisti}
\email{stefano.evangelisti@irsamc.ups-tlse.fr}
\affiliation{\lcpq}
\author{J.~A.~ Berger}
\email{arjan.berger@irsamc.ups-tlse.fr}
\affiliation{\lcpq}
\affiliation{\etsf}

\date{\today}

\begin{abstract}
\label{abstract}
We propose a simple direct-sum method for the efficient evaluation of lattice sums in periodic solids.
It consists of two main principles: i) the creation of a supercell that has the topology of a Clifford torus, which is a flat, finite and border-less manifold;
ii) the renormalization of the distance between two points on the Clifford torus by defining it as the Euclidean distance in the embedding space of the Clifford torus.
Our approach does not require any integral transformations nor any renormalization of the charges.
We illustrate our approach by applying it to the calculation of the Madelung constants of ionic crystals.
We show that the convergence towards the system of infinite size is monotonic, which allows for a straightforward extrapolation of the Madelung constant.
We are able to recover the Madelung constants with a remarkable accuracy, and at an almost negligible computational cost, i.e., a few seconds on a laptop computer.
\end{abstract}

\maketitle

\renewcommand{\baselinestretch}{1.5}
%
To describe properties of regular crystalline systems it is convenient to use periodic boundary conditions (PBC).
Indeed PBC are widely used in solid-state physics and chemistry as well as in material sciences.
In the case of short-range interactions PBC can be imposed via the Born - Von K\'arm\'an (BvK) boundary conditions.
However, often long-range interactions are present due to the Coulomb potential, and then BvK boundary conditions cannot be applied.
In this work we propose an alternative to the BvK boundary conditions that is compatible with the long-range Coulomb potential.

In our formalism we extract a supercell from the crystal, and then modify its topology into the topology of a torus.
However, in order to conserve the regular structure of the crystal, and in particular, the angles between the Bravais lattice vectors, we use a Clifford torus.
A Clifford torus is a flat, closed $d$-dimensional real Euclidean space embedded in a $d$-dimensional complex Euclidean space.
We note that, alternatively, the embedding space could be a $2d$-dimensional real space.
At the end, the supercell extracted from the original system has been transformed into
a toroidal manifold in which all the atoms that were equivalent in the original crystal are still equivalent.
We note that the use of flat tori has been suggested in the solid state, in order to impose PBC, in the so-called cyclic-cluster approach.
~\cite{thomas_bredow_development_2001}
Finally, a torus formalism has also been suggested by Mamode in order to compute the 
Madelung energy of hypercubic crystals of any dimension, through the solution of the Poisson
equation in a finite space.~\cite{mamode_fundamental_2014,mamode_computation_2017}

The topology of the Clifford torus raises the question how to define the distance between two points on the Clifford torus.
Naively, one might think to define this distance as the shortest distance \emph{on} the torus.
For example, for a one-dimensional supercell of length $L$ on a Clifford torus, i.e., a Clifford torus that is topologically equivalent to a circle with a circumference equal to $L$, such a distance between points $x_1$ and $x_2$ is given by $|x_1 - x_2|$ for $|x_1 - x_2| < L/2$ and $L - |x_1 - x_2|$ for $|x_1 - x_2| > L/2$.
However, the derivative of this distance with respect to $x_1$ or $x_2$ is discontinuous at $|x_1 - x_2| = L/2$.
Therefore, also the forces, which are related to the gradient of the Coulomb potential are discontinuous with respect to the position.
This is clearly an unphysical result.
However, there is another distance that is also naturally defined for the Clifford torus, namely the distance in the embedding space that contains the torus.
This distance is a unique and smooth function of the position and we will use it to define the Coulomb potential.
In summary, our approach consists of two main ideas: 1) We adapt the supercell to the topology of a Clifford torus ; 2) We renormalize the distance between two points on the Clifford torus as the distance between those points in the embedding space of the torus.~\cite{valenca_ferreira_de_aragao_simple_2019}

In this work we will illustrate our approach by applying it to the calculation of Madelung sums for the cohesion energy of ionic solids.
The calculation of these sums are conceptually difficult due to the fact that, because of the long-range nature of the Coulomb potential,
the resulting series is conditionally convergent.
For this reason, special care must be taken to perform such a summation, since the result depends on the order in which the summation is carried out.

Madelung sums are a special type of lattice sums which can be performed either by direct summations or indirectly by integral transformations.
In the case of direct summations it has been shown that neutrality of the supercell is required to accelerate convergence, and sometimes even to ensure convergence at all.~\cite{evjen_stability_1932} 
Therefore, fractional charges are sometimes needed in order to ensure convergence.
Two early methods of this type have been proposed by Evjen \cite{evjen_stability_1932} and by H{\o}jendahl \cite{k_hojendahl_no_1938}.
In Evjen's approach, for example, the surface charges are renormalized by applying weights equal to $1/8$, $1/4$, and $1/2$ for charges on the corners, edges, and faces, respectively. 
This method is numerically very efficient since it scales linearly with the number of atoms $N$.
Unfortunately, these approaches do not always converge.
Therefore, more advanced charge-renormalization techniques have been proposed~\cite{sousa_madelung_1993,derenzo_determining_2000,gelle_fast_2008}.
It has also been shown that the speed of convergence is directly related to the number of vanishing multipolar moments in the unit cell.~\cite{Marathe_PRB_1983, gelle_fast_2008}
In this way, by imposing a number of vanishing moments in the cell, exponential convergence can be achieved \cite{coogan_simple_1967,gelle_fast_2008}.
However, the main drawback of these approaches is that they require the renormalization of a large number of charges in the supercell.
Alternatively, a charge renormalization can also be achieved by adding neutralizing charges on the surface.~\cite{Wolf_1999,Pickard_2018}

The most commonly used integral-transformation method, on the other hand, was proposed by Ewald \cite{ewald_berechnung_1921},
and several related methods have been proposed in the literature.~\cite{nijboer_calculation_1957,glasser_lattice_1980,borwein_lattice_2013,borwein_convergence_1985,glasser_lattice_1980}
These approaches generally converge to the correct values.
However, the Ewald approach suffers from various drawbacks: 1) compared to most direct-sum approaches, its implementation is nontrivial; 2) it depends on an adjustable parameter that separates the short-range and long-range contributions and determines their rates of convergence; 3) in its standard implementation it scales as $N^2$ (with fast-Fourier transforms the scaling can be reduced to $N \log N$~\cite{Darden_1993}).
Therefore, a straighforward, accurate and numerically efficient direct-sum approach would be desirable.

We will show below that our approach based on a Clifford torus allows for a simple direct-sum approach yielding converged results.
No renormalization of the charges is required. Instead we use a simple renormalization of the distance between ions.
We note that, thanks to the PBC, all moments vanish by definition.
Moreover, our approach is parameter-free and scales linearly with $N$.

Let us consider a Bravais lattice in $d$ dimensions.
Let $\bv_j$ be the generator vectors of a unit cell 
(not necessarily a primitive unit cell).
A generic vector $|\bu\rangle$ belonging to the unit cell is given by
\begin{equation}
|\bu\rangle = \sum_{j=1}^d \alpha_j |\bv_j\rangle,
\end{equation}
with $0 \le \alpha_j < 1$.
Given a set of positive integers $K_1,\cdots,K_d$, 
we define the Euclidean supercell (ESC) as the parallelepiped generated by the vectors
\begin{equation}
|\bV_j\rangle = K_j |\bv_j\rangle.
\end{equation}
The ESC thus consists of $\prod_{j=1}^d K_j$ replicas of the unit cell.
A generic vector $|\bW^{ESC}\rangle$ in the ESC is given by
\begin{equation}
|\bW^{ESC}\rangle = |\bu\rangle + \sum_{j=1}^d k_j |\bv_j\rangle = \sum_{j=1}^d x_j |\bv_j\rangle
\end{equation}
where $x_j = \alpha_j + k_j$, with $0\le k_j \le K_j -1$.

In a completely analogous way, we define the {\em Clifford supercell} (CSC) as the Clifford torus associated to the ESC, obtained by joining the opposite edges of the corresponding ESC.
A generic point in the CSC is thus given by
\begin{equation}
|\bW^{CSC}\rangle = \sum_{j=1}^d
\frac{K_j}{2\pi} e^{i 2\pi x_j / K_j} |\bv_j\rangle,
\end{equation}
The factor $\frac{K_j}{2\pi}$ ensures that the circumference of a circle with such a radius coincides with the length of the corresponding edge of the ESC.
We remind the reader that a $d$-torus is the product of $d$ circles.
We note that, since the ESC and the corresponding CSC are built with the same unit vectors $\bv_j$, the two supercells are locally isometric.
\begin{figure}[t]
\centering
\includegraphics[width=\columnwidth]{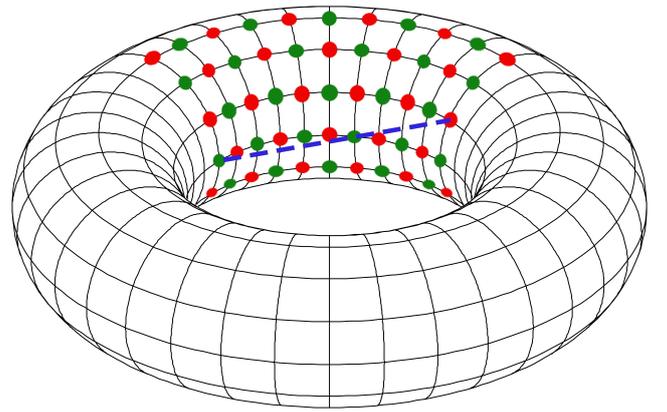}
\caption{An illustration of a Clifford supercell for a 2-dimensional NaCl structure; red dots represent Na$^+$ and green dots represent Cl$^-$.
The dashed blue line indicates the renormalized distance between two ions in the Coulomb potential. It is the shortest distance in the embedding space of the torus.
We note that a true Clifford torus has a flat surface which is impossible to represent graphically.}
\label{Fig:torus}
\end{figure}

To treat Coulomb potentials we have to define the distance between two points on the torus.
Since the CSC is embedded in $\mathbb{C}^d$, we define the distance between the two points $A$ and $B$ as the usual norm, in $\mathbb{C}^d$, 
of the difference $|{\bf R}_{AB}^{CSC}\rangle = |{\bf W}_B^{CSC}\rangle - |{\bf W}_A^{CSC}\rangle$ of the two corresponding position vectors,
\begin{equation}
|{\bf R}^{CSC}_{AB}\rangle = 
\sum_{j=1}^d\frac{K_j}{2\pi} \bigl(e^{i 2\pi x_j^B / K_j} - e^{i 2\pi x_j^A / K_j} \bigl) |\bv_j\rangle.
\end{equation}
Therefore, the distance $R_{AB}^{CSC} = \| \bR_{AB}^{CSC} \|$ between $A$ and $B$ is given by
\begin{equation}
R^{CSC}_{AB} = \left[\sum_{j=1}^d \frac{K_j^2}{2\pi^2}
\left[1-\cos\left(\frac{2\pi}{K_j}[x_j^B - x_j^A]\right)\right]\|{\bf v}_j\|^2\right]^{1/2},
\label{Eqn:dist_renorm}
\end{equation}
where for simplicity we assumed that $\langle \bv_i | \bv_j \rangle = \delta_{ij}$, since in practice one can often choose a supercell with orthogonal edges.
We note that this definition of the distance between two points is closely related to the modified position operator 
that we recently proposed for electrons in periodic systems.~\cite{valenca_ferreira_de_aragao_simple_2019}.
In Fig.~\ref{Fig:torus} we show an illustration of a CSC for a 2-dimensional NaCl structure and the renormalized distance between the ions.

Let us now consider the following general lattice sum
$S_A = \sum_{B}^{\prime} f \left(R^{CSC}_{AB}\right)$
where $B$ runs over the positions of all atoms in the CSC except that of the atom $A$ (indicated by the prime), and $f$ is an arbitrary real function.
More precisely, using Eq.~\eqref{Eqn:dist_renorm}, we can express this sum as
\begin{align}
S_A &= \sum_{B\in u.c.}\!\!\!\!{\vphantom{\sum}}' \,\, \sum_{k_1^B=0}^{K_1-1} \cdots \sum_{k_d^B=0}^{K_d-1}
f
\Bigg(\Bigg[\sum_{j=1}^d
\frac{K_j^2}{2\pi^2}
\bigg[1-
\notag
\\ &
\cos\left(\frac{2\pi}{K_j}[\alpha_j^B - \alpha_j^A + k_j^B - k_j^A]\right)\bigg]\|{\bf v}_j\|^2
\Bigg]^{1/2}
\Bigg),
\label{Eqn:doublesum}
\end{align}
where we used that $x_j = \alpha_j + k_j$.
The first summation on the right-hand side is over all the atoms $B$ in the unit cell except the atom $A$, which is also in the unit cell.
Thanks to the periodicity of the CSC and the substitution $k_j = k_j^B-k_j^A$ we can rewrite the above equation as
\begin{align}
S_A &= 
\sum_{B\in u.c.}\!\!\!\!{\vphantom{\sum}}' \,\,
\sum_{k_1=0}^{K_1-1} \cdots \sum_{k_d=0}^{K_d-1}
f
\Bigg(\Bigg[ \sum_{j=1}^d
\frac{K_j^2}{2\pi^2}
\bigg[1-
\notag
\\ &
\cos\left(\frac{2\pi}{K_j}[\alpha_j^B - \alpha_j^A + k_j]\right)\bigg]\|{\bf v}_j\|^2
\Bigg]^{1/2}\Bigg).
\label{Eqn:doublesum_simple}
\end{align}
%

The Madelung constant  $M_A$ is the electrostatic potential felt by an ion $A$ due to all other ions in the crystal.
It can be expressed as a lattice sum and its standard definition is given by
\begin{equation}
M_A = \sum_{B}\!{\vphantom{\sum}}' \frac{z_B}{R^{ESC}_{AB}/R_0},
\label{Eqn:Madelung_ESC}
\end{equation}
where $z_A$ is the valency of ion $A$, $R^{ESC}_{AB}$ is the distance between ions $A$ and $B$ in the ESC and $R_0$ is the nearest-neighbor distance.
Instead, in terms of the renormalized distance the Madelung constant of ion $A$ is redefined in our approach as
\begin{align}
M_A &= \sum_{B}\!{\vphantom{\sum}}' \frac{z_B}{R^{CSC}_{AB}/R_0}
\\ &=
R_0 \sum_{B\in u.c.}\!\!\!\!{\vphantom{\sum}}' z_B \sum_{k_1=0}^{K_1-1} \cdots \sum_{k_d=0}^{K_d-1}
\Bigg[ \sum_{j=1}^d
\frac{K_j^2}{2\pi^2}
\bigg[1-
\notag
\\ &
\cos\left(\frac{2\pi}{K_j}[\alpha_j^B - \alpha_j^A + k_j]\right)\bigg]\|{\bf v}_j\|^2
\Bigg]^{-1/2},
\label{Eqn:Madelung_CSC}
\end{align}
where we used Eq.~\eqref{Eqn:doublesum_simple}.
Equation \eqref{Eqn:Madelung_CSC} is the main result of this work.
We note that the cohesion energy $E_{coh}$ can be obtained from the knowledge of the Madelung constants of the ions in the lattice according to
\begin{equation}
E_{coh} = \frac{\sum_I z_I M_I}{R_0}.
\end{equation}
where the summation is over the various types of ions in the unit cell.

We now compare the following three approaches to calculate Madelung sums,
\begin{enumerate}
\item 
The plain sum over the ESC of increasing size according to Eq.~\eqref{Eqn:Madelung_ESC}.
This approach gives in general non-converging sums, and is added for completeness.
We note that the supercell is not electrically neutral in this approach.
\item
The sum over the ESC with surface-weighted charges according to Evjen's method, i.e.,
\begin{equation}
M_A = \sum_{B}\!{\vphantom{\sum}}' w_B \frac{z_B}{R^{ESC}_{AB}/R_0},
\end{equation}
where the weight $w_B$ is equal to $1/2$, $1/4$, and $1/8$ for an ion on the face, edge, and summit of the ESC, respectively; $w_B = 1$ for all ions inside the ESC.
\item
The sum over the CSC with renormalized distance according to Eq.~\eqref{Eqn:Madelung_CSC}, which is the method proposed in the present work.
\end{enumerate}
We limit ourselves to these three methods since they are of similar simplicity.
We note that since the numerical precision is a key issue in order to obtain a large number of significant digits, all our results have been obtained using quadruple precision.

To test our approach we first consider three types of cubic crystal structures, namely NaCl, CsCl, and ZnS, which represent, the rock-salt, the CsCl, and the zincblende structures, respectively.
In Tables \ref{Table:NaCl}, \ref{Table:CsCl}, and \ref{Table:ZnS}, we report the Madelung constants computed for a set of cubic supercells ($K=K_1=K_2=K_3$) 
of increasing size for NaCl, CsCl, and ZnS, respectively.
For simplicity $K$ will denote the number of unit cells per side in the following.
We see that, while the ESC and Evjen approaches only converge for NaCl, our CSC approach converges for all three structures.
More importantly, the CSC Madelung sums converge to the reference values.
Let us briefly discuss the three crystal structures in more detail.

\begin{figure}[t]
\centering
\includegraphics[width=\columnwidth]{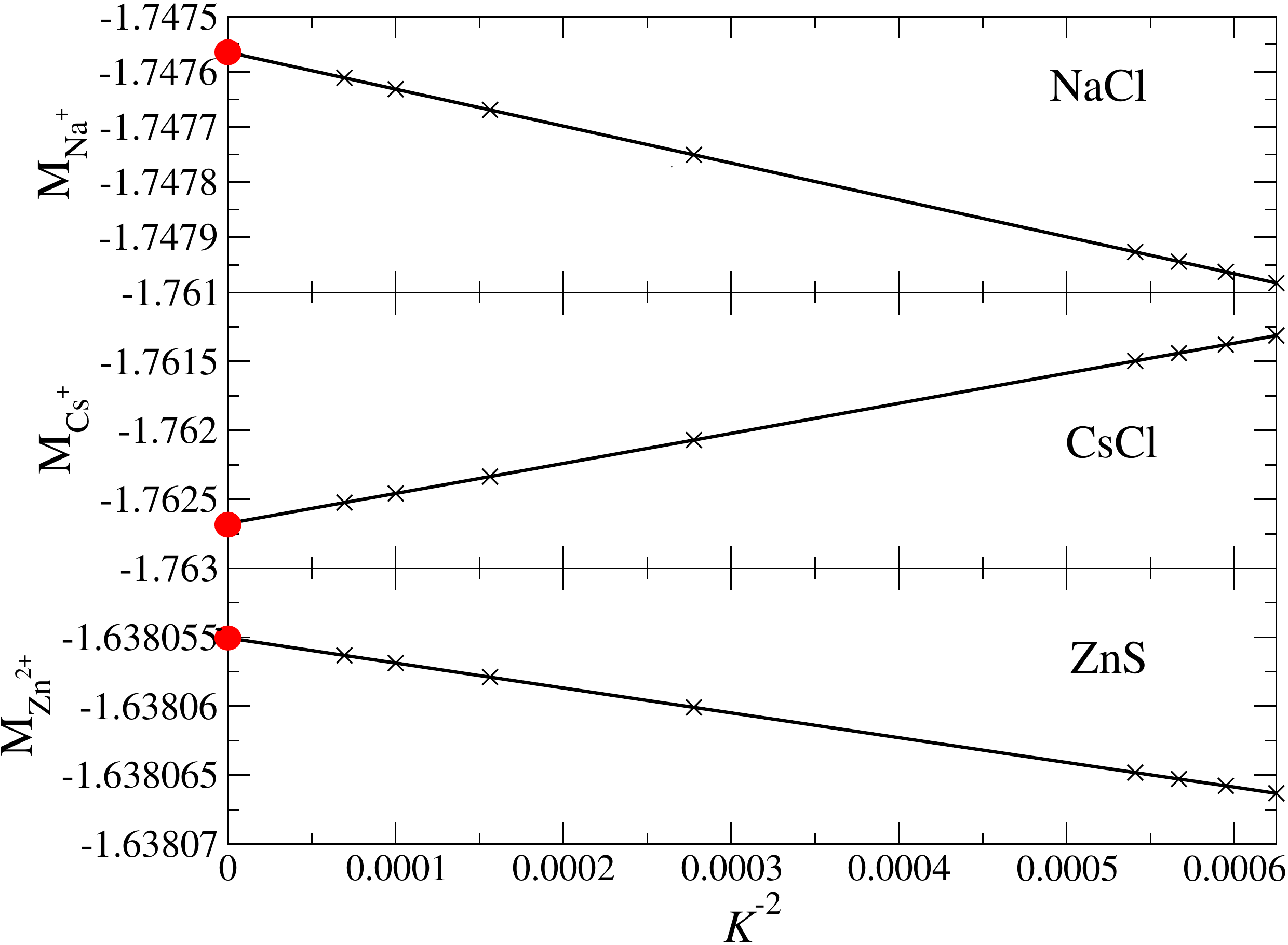}
\caption{The Madelung constants for the cations in NaCl, CsCl and ZnS, respectively, as a function of $K^{-2}$. The red dots indicate the extrapolated values which were obtained according to Eq.~\eqref{Eqn:extrapolation}.}
\label{Fig:extrapolation}
\end{figure}

\textit{NaCl}:
The Madelung constant for the NaCl crystal has been evaluated with high accuracy. 
Its fifteen-digits approximate value is $-1.74756459463318$~\cite{oeis_NaCl}.
The plain sum over ESC of increasing size converges extremely slowly to this limit with a series of alternating values that are above and below the exact value, 
depending on the total charge of the supercell.
In the case of NaCl the Evjen method converges extremely fast to the reference value.
Finally, the Clifford series converges to the reference value with a convergence rate in between that of the ESC and Evjen approaches.
As can be seen in Fig.~\ref{Fig:extrapolation} the convergence of the NaCl Madelung constant is monotonic in the CSC approach.
Thanks to this monotonicity we can extrapolate the finite-size results to that corresponding to the infinite-size CSC.
In Table \ref{Table:NaCl}, we also report this extrapolated CSC value, which coincides with the exact result up to the ninth decimal digit
using just 120 unit cells per side.
We will discuss the details of our extrapolation method below.

\begin{table}
\caption{The Madelung constant of $\mathrm{Na^+}$ in NaCl for various values of $K$, the number of unit cells per side.
The extrapolated $K \rightarrow \infty$ value has been obtained through a linear fit in $K^{-2}$ according to Eq.~\eqref{Eqn:extrapolation}
using the CSC results that correspond to the two largest $K$ values.}
\begin{center}
\begin{tabular}{lcccc}
\hline
\hline
$K$ & ESC &  Evjen &  CSC   \\
\hline
\\
40 &  -1.7333090325 & -1.7475646102 &   -1.7479830134   \\
41 &  -1.7614766492 & -1.7475645804 &   -1.7479628535   \\
42 &  -1.7339798824 & -1.7475646075 &   -1.7479441161   \\
43 &  -1.7608370145 & -1.7475645829 &   -1.7479266706   \\
60 &  -1.7380216149 & -1.7475645977 &   -1.7477505682   \\
80 &  -1.7403925416 & -1.7475645956 &   -1.7476692067   \\
100 & -1.7418198158 & -1.7475645950 &   -1.7476315469   \\
120 & -1.7427733060 & -1.7475645948 &   -1.7476110895   \\
\hline
$\infty$ & & & -1.7475645953 \\
\hline\hline
\end{tabular}
Reference value:~\cite{oeis_NaCl}   -1.7475645946
\end{center}
\label{Table:NaCl}
\end{table}

\textit{CsCl}:
The case of CsCl is well known, since by performing 
Evjen's approach one gets {\em two} different limits, i.e., one limit for $K$ even and another limit for $K$ odd.
The reason, as already discussed by Evjen himself in his paper, 
is because the surface of the supercell contains either only cations or only anions.
The commonly accepted value for the Madelung constant 
of this crystal is the average between the two limiting values, 
i.e., -1.76267477307098 \cite{oeis_CsCl}.
The values obtained with the plain ESC method are wildly oscillating.
Instead, our CSC approach is the only one that converges to the reference value.
Again, the convergence is monotonic (see Fig.~\ref{Fig:extrapolation}) and by extrapolating the values of the finite-size CSC 
we obtain a correspondence with the reference value up to the seventh decimal digit using just 120 unit cells per side.
\begin{table}
\caption{The Madelung constant of $\mathrm{Cs^+}$ in CsCl for various values of $K$, the number of unit cells per side.
The extrapolated $K \rightarrow \infty$ value has been obtained through a linear fit in $K^{-2}$ according to Eq.~\eqref{Eqn:extrapolation}
using the CSC results that correspond to the two largest $K$ values.}
\begin{center}
\begin{tabular}{lccc}
\hline
\hline
$K$ & ESC &  Evjen&  CSC   \\
\hline
40 &  -165.1951301706 &  -3.1228159774 & -1.7613129129  \\
41 &  -172.8428945898 &  -0.4025235314 & -1.7613786888  \\
42 &  -173.4399599212 &  -3.1228353436 & -1.7614398086  \\
43 &  -181.0877243486 &  -0.4025055166 & -1.7614967019  \\
60 &  -247.6434281092 &  -3.1229317065 & -1.7620703281  \\
80 &  -330.0917264008 &  -3.1229722138 & -1.7623349348  \\
100 & -412.5400247666 &  -3.1229909632 & -1.7624573245  \\
120 & -494.9883231553 &  -3.1230011482 & -1.7625237851  \\
\hline
$\infty$ & & & -1.7626748322 \\
\hline
\end{tabular}
Reference value:~\cite{oeis_CsCl}  -1.7626747731
\end{center}
\label{Table:CsCl}
\end{table}

\textit{ZnS}:
As in the previous case of the CsCl crystal, also for the ZnS crystal structure, only charges with the same sign are located on the faces of the ESC.
However, in this case Evjen's approach converges to the same incorrect limit for both even and odd numbers of unit cells on each side of the supercell.
We note that a different limit could be obtained if instead of an integer number of unit cells per side one would use a half-integer number of unit cells on each side of the supercell. The value for the Madelung constant of the ZnS crystal turns out to be the average of the two limiting values. It is given by -1.6380550533 \cite{oeis_ZnS}.
The sum over a plain ESC does not appear to converge, similarly to what
happens in the CsCl case, because the supercells are highly charged.
Again, our CSC method is the only one that converges to the reference bulk limit.
Moreover, the convergence is monotonic (see Fig.~\ref{Fig:extrapolation}) and by extrapolating the values of the finite-size CSC 
we obtain a correspondence with the reference value up to the seventh decimal digit using just 120 unit cells per side.

\begin{table}[b]
\caption{The Madelung constant of $\mathrm{Zn^{2+}}$ in ZnS for various values of $K$, the number of unit cells per side.
The extrapolated $K \rightarrow \infty$ value has been obtained 
through a linear fit in $K^{-2}$ according to Eq.~\eqref{Eqn:extrapolation} using the CSC results that correspond to the two largest $K$ values
(both corresponding to K even).}
\begin{center}
\begin{tabular}{lccc}
\hline
\hline
$K$&  ESC & Evjen &  CSC   \\
\hline
40 & 164.295318 & -2.3182037805 &    -1.6380663149 \\
41 &  168.405536  & -2.3182050224 &    -1.6380657779 \\
42 &  172.539858 & -2.3182062003 &    -1.6380652782 \\
43 & 176.650643 &  -2.3182072755 &    -1.6380648124 \\
60 & 246.741576 &  -2.3182182423 &    -1.6380600884 \\
80 & 329.188848 & -2.3182233050 &    -1.6380578914 \\
100 & 411.636528 & -2.3182256484 &    -1.6380568714 \\
120 & 494.084414 & -2.3182269215 &    -1.6380563166 \\
\hline
$\infty$ & & & -1.6380550555 \\
\hline\hline
\end{tabular}
Reference value: \cite{oeis_ZnS}  -1.6380550533
\end{center}
\label{Table:ZnS}
\end{table}

As can be seen from the tables above, the calculation of the Madelung constant from a single CSC, even of large size ($K\approx100$), 
yields an approximation of the exact value, corresponding to the system of infinite size, up to four or five digits.
An important advantage of our CSC approach is that we can obtain a much better approximation of the exact value by extrapolating the Madelung constants computed for the finite-size CSC's to the infinite-size limit.
It turns out that the CSC Madelung constants closely follows the following inverse power law as a function of the system size $K$,
\begin{equation}
M(K) = M_\infty + C K^{-2},
\label{Eqn:extrapolation}
\end{equation}
where $M_\infty$ is the Madelung constant of the infinite crystal and $C$ is a constant.
In Fig.~\ref{Fig:extrapolation} we used this inverse-power law to fit the computed Madelung constants of the finite systems.
We observe an almost perfect linearity of the curves for all three crystal structures.
Therefore, in practice, we used just the two largest values of $K$ in Eq.~\eqref{Eqn:extrapolation} to obtain $M_{\infty}$.
The extrapolated values in the tables were obtained in this way.
As can be seen from those results the inverse-power law yields very accurate results for $M_{\infty}$, 
increasing the correspondence with the reference values to seven, eight or even nine digits.

Finally, to demonstrate that our approach is not limited to cubic crystal structures with two types of ions 
we consider the two-dimensional hexagonal lattice with alternating positive and negative charges and the perovskite structure CaTiO$_3$.
The hexagonal lattice has an analytical solution~\cite{borwein_lattice_2013} to which we can compare.
For notational convenience we will refer to this structure as monolayer hexagonal BN ($h$-BN) despite the fact that the bonds in this material are also partly covalent.
We report the results in Table \ref{Table:extra}.
We observe that our CSC result for $h$-BN matches with the analytical reference value up to 10 digits.
For CaTiO$_3$ the most accurate reference value we found has only 4 digits.
Our CSC approach matches all 4 digits of the reference value.
\begin{table}[b]
\caption{The Madelung constants of $\mathrm{B^{+}}$ in monolayer $h$-BN and of $\mathrm{Ca^{2+}}$ in CaTiO$_3$.}
\begin{center}
\begin{tabular}{lcc}
\hline
\hline
 & CSC&  Reference   \\
\hline
$h$-BN  & -1.542219721703 & -1.542219721707 ~\cite{borwein_lattice_2013} \\
CaTiO$_3$ & -24.7549360589  & -24.7549~\cite{Lofgren_2014}  \\
\hline\hline
\end{tabular}
\end{center}
\label{Table:extra}
\end{table}

In conclusion, we presented a formalism suitable for the computation of lattice sums of ionic crystals.
The general strategy of our approach consists in transforming a supercell of a periodic system 
into a Clifford torus, and then renormalizing the distance between two
points on the torus as the Euclidean distance between these points in the embedding space of the torus.
In this way, a lattice sum on an infinite periodic system is replaced by a sequence of sums over finite periodic systems, 
and the value for the infinite crystal is then obtained by extrapolating to the infinite-size limit.
As a numerical illustration, we computed the Madelung constants of ionic crystals.
The values we obtain are in excellent agreement with the available reference data.
Moreover, our approach scales linearly with the number of atoms and is, therefore, numerically very efficient; the largest calculations performed in this work take a few seconds on a laptop computer.
Finally, we note that the same formalism can be applied to calculate the properties of electronic systems, e.g., Wigner crystals~\cite{Wigner,Diaz-Marquez_2018,escobar_azor_wigner_2019}.

This work was partially supported by the ``Programme Investissements d'Avenir'' under the
program ANR-11-IDEX-0002-02, reference ANR-10-LABX-0037-NEXT.
%
%
%
%
\end{document}